\documentclass[useAMS,usenatbib]{mn2e}
\usepackage[cp866]{inputenc}
\usepackage[english]{babel}
\usepackage{graphicx}
\usepackage{amsmath}

\def\la{\; \raise0.3ex\hbox{$<$\kern-0.75em\raise-1.1ex\hbox{$\sim$}}\;}
\def\ga{\;  \raise0.3ex\hbox{$>$\kern-0.75em\raise-1.1ex\hbox{$\sim$}}\;}
\def\pFn{p_{\raise-0.3ex\hbox{{\scriptsize F$\!$\raise-0.03ex\hbox{\rm n}}}}
}  
\def\pFp{p_{\raise-0.3ex\hbox{{\scriptsize F$\!$\raise-0.03ex\hbox{\rm p}}}}
}  
\def\pFe{p_{\raise-0.3ex\hbox{{\scriptsize F$\!$\raise-0.03ex\hbox{\rm e}}}}
}  
\def\pFmu{p_{\raise-0.3ex\hbox{{\scriptsize F$\!$\raise-0.03ex\hbox{\rm
$\mu$}}}} }  
\def\m@th{\mathsurround=0pt }
\def\eqalign#1{\null\,\vcenter{\openup1\jot \m@th
   \ialign{\strut$\displaystyle{##}$&$\displaystyle{{}##}$\hfil
   \crcr#1\crcr}}\,}

\newcommand{\dd}{\mbox{d}}                     

\title[Thermal evolution of a pulsating neutron star]
{Thermal evolution of a pulsating neutron star}
\author[M. E. Gusakov, D. G. Yakovlev, and O. Y. Gnedin]
{M. E. Gusakov$^{1, 2}$\thanks{E-mail:
gusakov@astro.ioffe.ru (MEG); yak@astro.ioffe.ru (DGY);
ognedin@astronomy.ohio-state.edu (OYG)}, 
 D. G. Yakovlev$^{1}$, and O. Y. Gnedin$^{3}$
\\
$^{1}$Ioffe Physical Technical Institute, 
Politekhnicheskaya 26, St.-Petersburg 194021, Russia \\
$^{2}$N. Copernicus Astronomical Center, Bartycka 18,
Warsaw 00-716, Poland \\
$^{3}$Ohio State University, 760 1/2 Park Street, 
Columbus, OH 43215, USA}
\begin{document}

\date{Accepted 2005 xxxx. Received 2005 xxxx; 
in original form 2005 xxxx}

\pagerange{\pageref{firstpage}--\pageref{lastpage}} \pubyear{2005}

\maketitle

\label{firstpage}

\begin{abstract}
We have derived a set of equations 
to describe the thermal evolution of a neutron 
star which undergoes small-amplitude 
radial pulsations. 
We have taken into account,
in the frame of the General Theory of Relativity, the pulsation 
damping due to the bulk and shear 
viscosity and the accompanying heating of 
the star. 
The neutrino emission of 
a pulsating non-superfluid star and 
its heating due to the bulk viscosity 
are calculated assuming 
that both processes are determined by  
the non-equilibrium modified Urca process. 
Analytical and numerical solutions to the set of 
equations of the stellar evolution 
are obtained for linear and strongly non-linear deviations
from beta-equilibrium. 
It is shown that a 
pulsating star may be heated to very 
high temperatures, while the pulsations 
damp very slowly with time 
as long as the damping is determined by the 
bulk viscosity
(a power law damping during 100--1000 years). 
The contribution of the shear 
viscosity to the damping becomes important 
in a rather cool star with a low 
pulsation energy.
\end{abstract}


\section{Introduction}
\label{introduction}

Dissipation processes play an important role 
in the neutron star physics; for instance, 
they determine the 
damping of stellar pulsations 
(see, e.g., Cutler, Lindblom $\&$ Splinter 1990). 
Pulsations may be excited during the star 
formation or during its evolution under the action 
of external perturbations or internal instabilities. 
The instabilities arising in a 
rotating star from the emission of 
gravitational waves may be suppressed by 
dissipation processes. 
This affects the maximum rotation frequency 
of neutron stars and creates 
problems in the detection 
of gravitational waves 
(see, e.g., Zdunik 1996; Lindblom 2001; 
Andersson $\&$ Kokkotas 2001; Arras et al.\ 2003).

The joint thermal and pulsational evolution of
neutron stars was studied long ago (e.g., Finzi \& Wolf
1968 and references therein). Naturally, it was
done with a simplified physics input and under
restricted conditions (Section \ref{system}).
However later, while estimating the characteristic 
times of pulsational damping, one 
usually ignored the temporal evolution of 
the stellar temperature 
(see, e.g., Cutler $\&$ Lindblom 1987; Cutler et al.\ 1990), 
which led to an exponential 
damping. This 
is not always justified because the parameters defining 
the damping rate, e.g., 
the bulk and shear viscosity coefficients, 
are themselves temperature-dependent. 

Clearly, the temperature 
variation can be neglected  
if the characteristic damping time 
$\tau \ll t_{\rm cool}$ 
and $E_{\rm puls} \ll E_{\rm th}$, 
where $t_{\rm cool}$ is 
the characteristic time of neutron star cooling, 
while $E_{\rm puls}$ and $E_{\rm th}$ 
are the pulsational and thermal 
energies, respectively. We will show that these 
conditions are violated in a wide range of initial 
temperatures and pulsation amplitudes.

This paper presents a self-consistent 
calculation of the dissipation of radial pulsations 
with account for the thermal 
evolution of a non-superfluid neutron star 
whose core consists of neutrons (n), 
protons (p) and electrons (e). 
It extends the consideration by 
Finzi \& Wolf (1968) (see Section \ref{system} for details). We consider two  
dissipation mechanisms: one is via the {\it non-linear}
(in the pulsation amplitude)
bulk viscosity in the stellar core and the 
other is due to the shear viscosity. We neglect other possible 
dissipation mechanisms, 
particularly, the 
damping of pulsations induced by the 
star magnetic field (as discussed in detail by 
McDermott et al.\ 1984 and by 
McDermott, van Horn $\&$ Hansen 1988). 
The magnetic field is assumed to be low.

\section[]{Eigenfunctions and Eigenfrequencies of 
non-dissipative Radial Pulsations}
\label{radial}
Here we discuss briefly radial pulsations 
of a neutron star, ignoring 
energy dissipation. 
This problem was first considered 
by Chandrasekhar (1964), and we will 
refer to his results. 
The metric for a spherically symmetric star, 
which experiences radial pulsations, can be written as  
\begin{equation}
 \dd s^2 =  -{\rm e}^{\nu} \dd t^2 + r^2 \dd \Omega^2
    + {\rm e}^{\lambda} \, \dd r^2,
\label{ds}
\end{equation}
where $r$ and $t$ are the radial and time coordinates and 
$\dd \Omega$ is a solid angle element 
in a spherical frame with the origin at 
the stellar center. Here and below, we use 
the system of units, in which light velocity 
$c$ = 1. The 
functions $\nu$ and $\lambda$ 
depend only on $r$ and $t$ and can be written as:
$\nu(r,t)=\nu_0(r)+\delta \nu(r,t)$;
$\lambda(r,t)=\lambda_0(r)+\delta \lambda(r,t)$.
Here $\nu_0(r)$ and
$\lambda_0(r)$ are the metric functions for an unperturbed (equilibrium)
star, and $\delta \nu(r,t)$ and $\delta \lambda(r,t)$
are the metric perturbations due to the radial pulsations 
(described by Eqs.\ (36) and (40) of Chandrasekhar 1964).

The radial pulsations  
can be found by solving the Sturm-Liouville
problem (Eq.\ (59) of Chandrasekhar 1964). 
A solution gives  
eigenfrequencies of pulsations $\omega_k$ 
and eigenfunctions $\xi_k(r)$, 
where $\xi_k(r)$ is the 
Lagrangian displacement of a fluid element with 
a radial coordinate $r$. 
By neglecting the 
dissipation and the non-linear interaction between 
the modes (the pulsation 
amplitude is taken to be small, $|\xi_k(r)| \ll r$), 
we can write the general solution for 
a $k$-th mode as $\xi(r,t)=\xi_k(r) \cos \omega_k t$. 
The boundary conditions for the Sturm-Liouville problem have 
the form: $P \left(r=R + \xi(R, t) \right)=0$, $\xi(0, t)=0$, 
where $P(r,t)$ is the pressure  
and $R$ is the unperturbed stellar radius.  
\begin{figure*}
\setlength{\unitlength}{1mm}
\leavevmode
\hskip  0mm
\includegraphics[width=120mm,bb=18  145  554  690,clip]{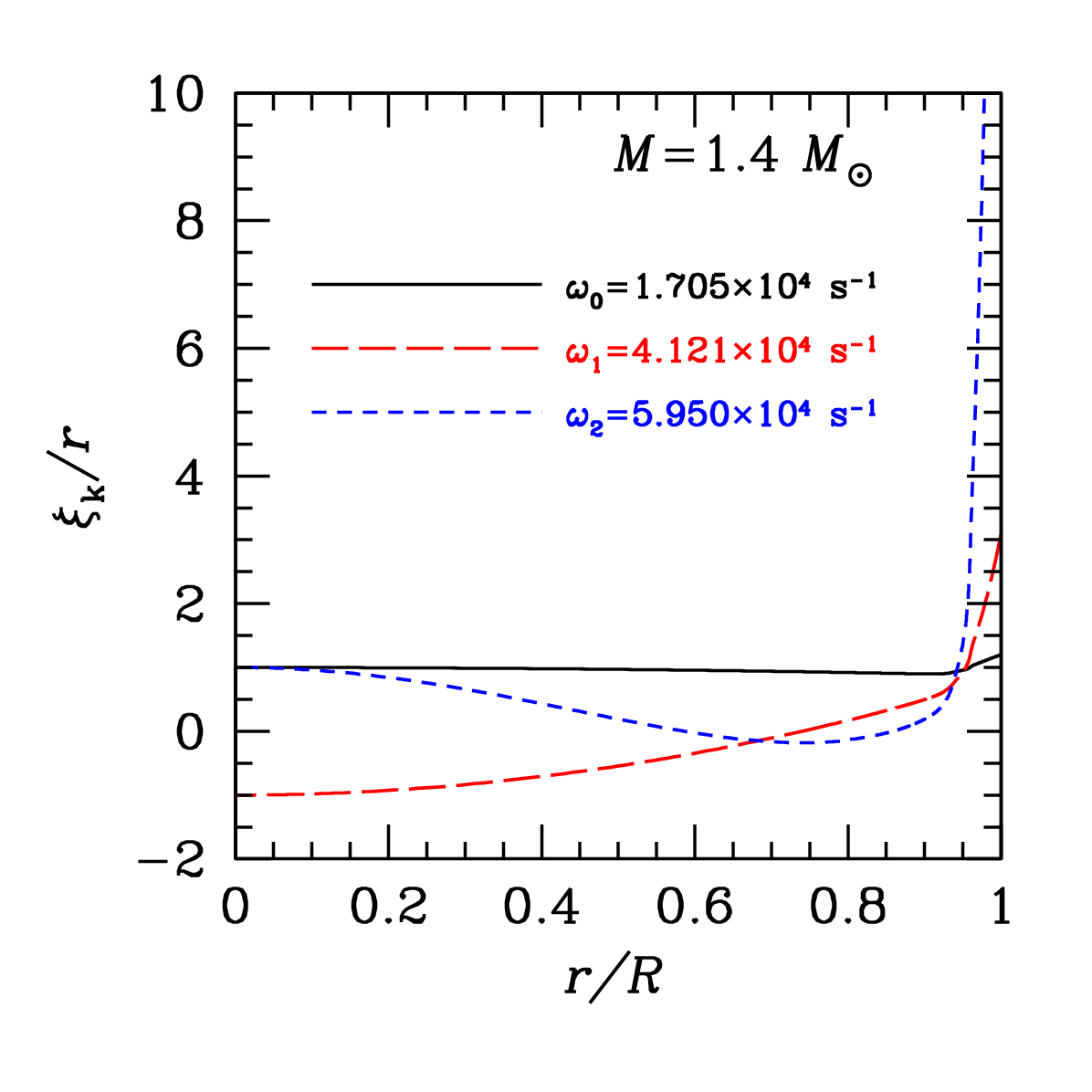}
\caption
{
The parameter $\xi_k/r$ normalized such that $|\varepsilon|=1$,
for the fundamental, first, 
and second modes of
radial stellar pulsations (solid, long-dashed, and short-dashed lines, 
respectively) versus the dimensionless
radial coordinate $r/R$.
}
\label{1}   
\end{figure*}

We employ the equation of state 
of Negele \& Vautherin (1973) in the stellar 
crust and the equation of state of
Heiselberg \& Hjorth-Jensen (1999) in the stellar core. 
The latter equation of state 
is a convenient analytical approximation 
of the equation of state of Akmal \& Pandharipande (1997).
For this equation of state,
the most massive stable neutron 
star has the central density
$\rho_{\rm c}=2.76 \times 10^{15}$ g~cm$^{-3}$,
the circumferential radius $R=10.3$ km, and 
the mass $M = M_{\rm max} = 1.92 \, M_\odot$.
The powerful direct Urca process 
of neutrino emission is open in the 
core of a star of mass $M > 1.83\, M_\odot$.

An important parameter which enters 
the equation of radial pulsations is the 
adiabatic index $\gamma$. 
Since the frequency of stellar pulsations
is $\omega_k \gg 1/t_{\rm Urca}$, 
where $t_{\rm Urca}$ is the 
characteristic beta-equilibration time 
(see, e.g., Haensel, Levenfish $\&$ Yakovlev 2001; Yakovlev et al.\ 2001), 
the adiabatic index must be determined assuming 
the ``frozen'' nuclear composition  
(see, e.g., Bardeen, Thorne $\&$ Meltzer 1966):
\begin{equation}
      \gamma = \frac{\partial \ln P(n_{\rm b}, x_{\rm e})}
              {\partial \ln n_{\rm b}},
\label{gamma}
\end{equation}
where $n_{\rm b}$ is the baryon number density, 
$x_{\rm e} = n_{\rm e}/n_{\rm b}$, 
and $n_{\rm e}$ is the electron number density.

The relative radial displacement of 
matter elements in a pulsating star 
(in the absence of dissipation effects) 
will be described by a small parameter $\varepsilon$:
\begin{equation}
       \varepsilon = \lim_{r \rightarrow 0} \, \xi_k(r)/r.
\label{CC}
\end{equation}
Thus, $\varepsilon$ determines the 
normalization of the function $\xi_k(r)$.

Figure 1 shows the dependence of $\xi_k(r)/r$ 
(artificially normalized such that $|\varepsilon|=1$) 
on the distance to the stellar center $r$ 
for the first three 
modes with the frequencies 
$\omega_0=1.705 \times 10^4$ s$^{-1}$ 
(solid line), 
$\omega_1=4.121 \times 10^4$ s$^{-1}$ (long dashed line), 
and $\omega_2=5.950 \times 10^4$ s$^{-1}$ 
(short dashed line), respectively. 
By way of illustration, we consider a model of a star 
of mass $M = 1.4 \, M_\odot$
($R=12.17$ km, $\rho_{\rm c}=9.26 \times 10^{14}$ g~cm$^{-3}$). 
As expected,
the fundamental mode is close to the 
homological solution $\xi_0(r) = r$. 
Introducing a  
normalization constant, we get $\xi_0(r)= \varepsilon\,r$.
Therefore, for the fundamental mode, $\varepsilon$ 
determines the amplitude of relative displacements 
of the pulsating stellar surface.

We will further need the pulsation energy, 
which can be calculated if we
formulate, for example, 
the variational principle for 
the characteristic eigenvalue problem in question.
For the $k$-th radial mode, we have 
(see, e.g., Meltzer $\&$ Thorne 1966)
\begin{equation}
E_{\rm puls} = \frac{1}{2} \,\, \int \, (P+\rho)
       \left[ {\rm e}^{(\lambda_0-\nu_0)/2} \omega_{\rm k}
       \, \xi_{\rm k}\right]^2
       {\rm e}^{\nu_0/2} \, \dd V \,,
\label{Epuls}
\end{equation}
where $\rho$ is the mass density and 
$\dd V = 4 \pi r^2 {\rm e}^{\lambda_0/2} \, \dd r$
is the volume element measured in a comoving frame.
	
The account of the energy dissipation in the 
$k$-th mode leads to a relatively slow damping of pulsations. 
In particular, we take for the Lagrangian displacement 
\begin{equation}
      \xi(r,t) = C_k(t) \, \xi_k(r) \, \cos \omega_k t.
\label{Ck}
\end{equation}
where $C_k(t)$ is a slowly decreasing function 
of time (the characteristic dissipation 
time $\tau \gg 1/\omega_k$), 
which will be further termed {\it the pulsation amplitude}. 
The dissipation is assumed to be ``switched on'' 
at the moment of time $t=0$, at which the 
initial amplitude is
\begin{equation}
   C_k(0) = 1.
\label{start}
\end{equation}
From Eq.\ (\ref{Epuls}) the pulsation energy 
in the $k$-th mode with dissipation is        	     
\begin{equation}
         E_{\rm puls}(t) = E_{\rm puls 0} \, C_k^2(t).
\label{Epuls1}
\end{equation}
Using Eqs.\ (\ref{Epuls}) and (\ref{Epuls1}), 
we can estimate the pulsation energy for the fundamental mode, 
$E_{\rm puls}(t)
\sim 2 \times 10^{53} \, \omega^2_4 \, \varepsilon^2 \, C_k^2(t)$ erg.
The thermal energy of the star is 
$E_{\rm th} \sim (4 \pi /3) \, R^3 \,
c_T \, T \sim 10^{48} \, T_9^2$ erg,
where $c_T \propto T$
is the specific (per unit volume) heat capacity 
of the stellar matter 
(see, e.g., Yakovlev, Levenfish $\&$ Shibanov 1999)
and $T_9$ is the internal temperature of the star
in units of $10^9$ K, 
$\omega_4=\omega_k/(10^4 \,$ {\rm s}$^{-1}$).
These estimates show that there is 
a wide range of values of the parameters 
$\varepsilon$, $C_k(t)$, and $T$,
at which $E_{\rm puls} \ga E_{\rm th}$. 
In such a case, one should account, at least, for 
the stellar temperature evolution during 
the damping of pulsations.

%
\section{Non-equilibrium Modified Urca Process}
\label{noneq}
%
The condition for beta-equilibrium in the stellar core has the form:
$\delta \mu(r,t) = \mu_{\rm n}-\mu_{\rm p}-\mu_{\rm e} = 0$,
where $\mu_i$ is the chemical potential of 
particle species $i={\rm n, p, e}$. 
Stellar pulsations lead to deviations from 
beta-equilibrium, 
$\delta \mu(r,t) \ne 0$,
and hence, to the 
dissipation of the pulsation energy. 
The dissipation rate is determined by the 
processes tending to return the system to equilibrium. 
We suggest that the direct 
Urca process in the neutron star core is forbidden. 
Then the main process which determines the 
pulsation energy dissipation is the modified Urca process. 
In this section, we will 
discuss the non-equilibrium modified Urca process 
and obtain the relationship 
between the Lagrangian displacement 
$\xi(r,t)$
and the parameter $\delta \mu(r,t)$
that characterizes the local deviation from beta-equilibrium.

The non-equilibrium modified Urca process has been discussed 
since the end of the 1960s (see the classical
paper by Finzi \& Wolf 1968 and references therein).
However those old studies were accurate only
qualitatively (see, e.g., Haensel 1992 and Section \ref{system}).
Later the problem has been reconsidered by several 
authors (see, e.g., Haensel 1992; 
Reisenegger 1995; Haensel et al.\ 2001). 
Below, references will primarily be made to the review of 
Yakovlev et al.\ (2001) since we employ similar notations. 
The modified Urca process has the neutron 
and the proton branch, each including a direct and an 
inverse reaction:
\begin{equation}
 {\rm n + N} \to {\rm p} + {\rm N} + {\rm e} + \bar{\nu}_{\rm e} \, , 
 \quad
 {\rm p + N + e} \to {\rm n + N}  + \nu_{\rm e}.
\label{murcan}
\end{equation}
Here ${\rm N=n}$ or ${\rm p}$
for the neutron or proton branch, respectively. 
In beta-equilibrium, the neutrino emissivities 
in these two channels are given by
\begin{eqnarray}
     Q_{\rm eq}^{\rm (n)} & \approx & 8.1 \times 10^{21}
                \left( {m_{\rm n}^\ast \over m_{\rm n} } \right)^3
                \left( {m_{\rm p}^\ast \over m_{\rm p} } \right)
\nonumber \\                
&\times&        \left( {n_{\rm p}  \over n_0 } \right)^{1/3} \!
                T_9^8 \, \alpha_{\rm n} \beta_{\rm n} \; \; \;
                {\rm erg \; cm^{-3} \; s^{-1}}  \, ,
\label{Qn} \\
     Q_{\rm eq}^{\rm (p)} & \approx & Q_{\rm eq}^{\rm (n)} \, \, \,
       \left( m_{\rm p}^\ast \over m_{\rm n}^\ast \right)^2
      {(\pFe+3 \pFp - \pFn)^2 \over
      8 \pFe \pFp} \,
    \Theta,
\label{Qp}
\end{eqnarray}
where $n_0=0.16$ fm$^{-3}$ 
is the nucleon number density in atomic nuclei; 
$n_{\rm p}$ is the 
proton number density; 
$m_{\rm n}$ and $m_{\rm p}$ 
are the masses of free neutrons and protons; 
$m_{\rm n}^\ast$ and $m_{\rm p}^\ast$
are the effective masses of neutrons and protons in dense matter; 
$\pFe, \pFp$, and $\pFn$
are, respectively, the Fermi momenta of electrons, 
protons, and neutrons; 
and $\alpha_{\rm n}, \beta_{\rm n} \sim 1 $
are the correction factors
(for details, see Yakovlev et al.\ 2001). 
In Eq.\ (\ref{Qp}) the 
function $\Theta$=1 
if the proton branch is allowed by momentum conservation
($\pFn < 3 \pFp+\pFe$), and $\Theta=0$ otherwise.

In beta-equilibrium, the direct 
and inverse reaction rates in Eq.\ (\ref{murcan})
coincide, i.e., the matter composition 
does not change with time. The reactions 
involve only particles in the vicinity 
of their Fermi surfaces, with the energy 
$|\epsilon_i - \mu_i| \la k_{\rm B} T$, 
where 
$i={\rm n,p,e}$, and $k_{\rm B}$
is the Boltzmann constant. 
Therefore, the neutrino 
emissivity depends sensitively on the temperature, 
and the process cannot occur at $T=0$. 
A drastically different situation arises 
in the presence of deviations from  
beta-equilibrium ($\delta \mu \ne 0$). 
The rates of the direct and inverse 
reactions become different, the 
system tends to equilibrium, 
and the matter composition changes; the process 
remains open even at $T=0$.
	
Let $\Gamma$ and $\bar{\Gamma}$ 
be the numbers of direct and inverse reactions 
of the modified 
Urca process per unit volume per unit time. 
The analytical expressions for
$\Delta \Gamma = \bar{\Gamma}-\Gamma$ 
and for the neutrino emissivity 
$Q_{\rm noneq}=Q^{\rm (n)}+Q^{\rm (p)}$ 
of the non-equilibrium modified Urca process 
were derived by Reisenegger (1995):
\begin{eqnarray}
      \Delta \Gamma  &=&
        {14680 \over 11513 
	}
        \, {Q_{\rm eq} \over k_{\rm B} T}\, y \,H(y) ,
\label{dG} \\
       Q_{\rm noneq}  &=&  Q_{\rm eq} \,F(y),
\label{Q}
\end{eqnarray}
where 
$Q_{\rm eq} = Q_{\rm eq}^{\rm (n)}+Q_{\rm eq}^{\rm (p)}$;
and the functions $H(y)$ and $F(y)$ are given by
\begin{eqnarray}
      H(y) &  = &
                  1+ {189 \pi^2\, y^2   \over 367}
                     + { 21  \pi^4\, y^4 \over 367}
                     + { 3  \pi^6\, y^6 \over 1835},
\label{HH} \\
      F(y) & = & 1 + {22020 \pi^2\, y^2 \over 11513}
                   + {5670 \pi^4 \, y^4 \over 11513 }
\nonumber \\
&&+                  {420 \pi^6\, y^6 \over 11513}
                   + {9 \pi^8\, y^8 \over 11513 }.
\label{FF}
\end{eqnarray}
Here 
$y \equiv \delta \mu/ (\pi^2 k_{\rm B} T)$;
the factor $\pi^2$ 
in the denominator is introduced to emphasize 
that the real variation scale 
of the functions  $H(y)$ and $F(y)$ 
is $\delta \mu/(10 k_{\rm B} T)$
(but not just $\delta \mu/k_{\rm B} T$). 
It follows from Eqs. (\ref{dG}) and (\ref{Q}) 
that there are two pulsation regimes. 
The regime with 
$\delta \mu \ll k_{\rm B}T$ ($y \ll 1$)
will be 
referred to as {\it subthermal} 
and that with 
$\delta \mu \gg k_{\rm B} T $ ($y \gg 1$)
as {\it suprathermal}. Relative 
displacements of fluid elements in both regimes 
are taken to be small ($\varepsilon \ll 1$). 
From these equations one can see that the values of 
$\delta \mu \, \Delta \Gamma$ and $Q_{\rm noneq}$
in the suprathermal regime
are independent of temperature.

Let us now find the relationship between 
the Lagrangian displacement $\xi(r,t)$
and the chemical potential difference $\delta \mu(r,t)$.
The quantity $\delta \mu$ can be 
treated as a function of three thermodynamic variables, 
say, $n_{\rm b}, n_{\rm e}$ and $T$: 
$\delta \mu = \delta \mu(n_{\rm b}, n_{\rm e}, T)$. 
During pulsations, these variables will deviate from their 
equilibrium values 
$n_{\rm b0}, \, n_{\rm e0}$ and $T_0$ 
by $\Delta n_{\rm b}(r,t), \,
\Delta n_{\rm e}(r,t)$ and $\Delta T(r,t)$.
Taking the deviations to 
be small, i.e. obeying the 
inequality $\varepsilon \ll 1$, one can write:                  
\begin{align}
       &\delta \mu(r,t) = {\partial \delta \mu(n_{\rm b0}, n_{\rm e0}, T_0)
       \over \partial n_{\rm b0}}\, \Delta n_{\rm b}(r,t)
\nonumber \\       
&+    {\partial \delta \mu(n_{\rm b0}, n_{\rm e0}, T_0)
       \over \partial n_{\rm e0}}\, \Delta n_{\rm e}(r,t)+
       {\partial \delta \mu(n_{\rm b0}, n_{\rm e0}, T_0)
       \over \partial T_{0}}\, \Delta T(r,t).
\label{series}
\end{align}
The last term in Eq.\ (\ref{series}) 
can be neglected because
$\partial \delta \mu(n_{\rm b0}, n_{\rm e0}, T_0)
/ \partial T_{0} \propto T_0$ and  
$\Delta T(r,t) \sim \Delta n_{\rm b}(r,t) \, T_0/n_{\rm b0}$ 
(see, e.g., Reisenegger 1995).
Accordingly, for a 
strongly degenerate matter 
($\mu_i \gg k_{\rm B} T_0$, $i={\rm n,p,e}$),
this term is much smaller 
than the first two terms. 
The temperature $T$ will further 
denote an ``average'' temperature $T_0$ 
and its oscillations around 
the equilibrium value will be neglected. 
	
The form of the functions 
$n_{\rm b}(r,t)$ and $n_{\rm e}(r,t)$ 
can be found from the continuity 
equations for baryons and electrons:
\begin{eqnarray}
    (n_{\rm b}u^\alpha )_{; \, \alpha}&=& 0,
\label{consnb} \\
     (n_{\rm e}u^\alpha )_{; \, \alpha}&=&\Delta \Gamma.
\label{consne}
\end{eqnarray}
Here $u^\alpha=\dd x^\alpha/{\rm \dd s}$
is the velocity four-vector of the pulsating matter. 
Note that the source $\Delta \Gamma$
in the continuity equation for electrons 
is responsible for beta-relaxation processes.
	
Writing explicitly the covariant derivatives in 
Eqs.\ (\ref{consnb}) and ({\ref{consne}}) in 
the metric ({\ref{ds}}) and 
neglecting all terms which are
quadratic and higher order in $\xi(r,t)$, one obtains:
\begin{eqnarray}
  {\partial n_{\rm b} \over \partial t} + {{\rm e}^{\nu_0/2} \over {r^2}} \,
  {\partial \over \partial r} \left(  n_{\rm b0} r^2 {\rm e}^{-\nu_0/2} \,
  {\partial \xi(r,t) \over \partial t} \right) &=& 0,
\label{consnb_exp} \\
  {\partial n_{\rm e} \over \partial t} + {{\rm e}^{\nu_0/2} \over {r^2}} \,
  {\partial \over \partial r} \left(  n_{\rm e0} r^2 {\rm e}^{-\nu_0/2} \,
  {\partial \xi(r,t) \over \partial t} \right)
  &=& \Delta \Gamma  {\rm e}^{\nu_0/2}.
\label{consne_exp}
\end{eqnarray}
These expressions have been derived using Eq.\ (36) of
Chandrasekhar (1964) for 
the correction $\delta \lambda(r,t)$ 
to the metric (see Section 2):
\begin{equation}
\delta \lambda(r,t) = - \xi(r,t) \, {\dd \over \dd r} (\lambda_0+\nu_0).
\label{lambda}
\end{equation}
Equation (\ref{consnb_exp}) is easily integrated and yields
\begin{eqnarray}
   \Delta n_{\rm b}(r,t)
   &\equiv& n_{\rm b}(r,t) - n_{\rm b0}
   \nonumber \\
    &=& - {{\rm e}^{\nu_0/2} \over r^2} \, {\partial \over \partial r}
   \left( n_{\rm b0} r^2 {\rm e}^{-\nu_0 /2} \, \xi(r,t) \right).
\label{nbnb}
\end{eqnarray}
The solution to Eq.\ (\ref{consne_exp}) can be written as
\begin{eqnarray}
   \Delta n_{\rm e}(r,t) & \equiv & n_{\rm e}(r,t) - n_{\rm e0} =
   \Delta n_{\rm e0}(r,t) + \Delta n_{\rm e1}(r,t),
\label{nene} \\
   \Delta n_{\rm e0}(r,t) &=&-{{\rm e}^{\nu_0/2} \over r^2} \,
   {\partial \over \partial r}
   \left( n_{\rm e0} r^2 {\rm e}^{-\nu_0 /2} \, \xi(r,t) \right) \,,
\label{nene0}
\end{eqnarray}
where $\Delta n_{\rm e0}(r,t)$
describes variations of the electron number density  
ignoring beta-processes, 
while the function
$ \Delta n_{\rm e1}(r,t)$ 
describes variations determined by these 
processes. The latter function satisfies the equation
\begin{equation}
    {\partial \Delta n_{\rm e1} \over \partial t} =
    \Delta \Gamma \, {\rm e}^{\nu_0/2} \,.
\label{ne1}
\end{equation}
Generally, the source $\Delta \Gamma$ 
is a complicated function of the electron 
number density $n_{\rm e}(r,t)$. 
We are, however, 
interested in the high frequency limit, where
$\omega_k \gg 1/t_{\rm Urca}$ 
(see Section 2).
In that case, the source in 
the right-hand side of Eq.\ (\ref{consne_exp}) is 
smaller than other terms. 
This means that changes in the electron 
number density due to beta-transformations 
are relatively small in a pulsating 
star (see, e.g., Haensel et al.\ 2001). 
Therefore, the small parameter $\Delta n_{\rm e1}$ 
can be omitted in Eq.\ (\ref{series}).
	
By substituting the expressions for 
$\Delta n_{\rm b}$ and $\Delta n_{\rm e}$ 
from Eqs.\ (\ref{nbnb}) and (\ref{nene0}) into 
Eq.\ (\ref{series}), 
we find the relationship between
$\delta \mu(r,t)$ and $\xi(r,t)$:
\begin{equation}
   \delta \mu(r,t) = - {\partial \delta \mu(n_{\rm b0}, x_{\rm e0})
   \over \partial n_{\rm b0}} \, \,
   n_{\rm b0} \, {{\rm e}^{\nu_0/2} \over {r^2}} \,
   {\partial \over \partial r} \left( r^2 {\rm e}^{-\nu_0/2} \,
   \xi(r,t) \right).
\label{connection}
\end{equation}
Note that the partial derivative with respect to $n_{\rm b0}$ 
is taken at constant  
$x_{\rm e0}=n_{\rm e0}/n_{\rm b0}$. 
Using Eq.\ (\ref{connection}), 
we can express the parameter 
$y=\delta \mu/(\pi^2 k_{\rm B} T)$, 
as well as $\Delta \Gamma$ and $Q_{\rm noneq}$
(see Eqs. (\ref{dG}) and (\ref{Q})), 
through the Lagrangian displacement $\xi(r,t)$. 
The relationship 
between $\delta \mu(r,t)$ and $\xi(r,t)$ for non-radial pulsations
can be derived in a similar way.

\section{The Equations of Stellar Thermal Evolution and 
Pulsation Damping out of Beta-Equilibrium}
\label{system}

The thermal balance equation for 
a pulsating neutron star will be derived taking into account 
three dissipation mechanisms: 
the shear viscosity in the core, 
the non-equilibrium beta-processes in the core, 
and heat conduction. 
The equations 
of relativistic fluid dynamics to describe 
energy-momentum conservation are written as
\begin{equation}
    T^{\alpha \beta}_{\, ; \beta} = - Q_\nu \, u^\alpha \,,
\label{Tab}
\end{equation}
where $Q_\nu$ is the total neutrino emissivity 
of all processes 
(including the non-equilibrium modified Urca process 
described by Eq.\ (\ref{Q})); 
$T^{\alpha \beta}$ is the energy-momentum 
tensor to be written as (see, e.g., Weinberg 1971):
\begin{eqnarray}
       T^{\alpha \beta} &=& P g^{\alpha \beta} + (P+\rho) \,
       u^{\alpha} u^{\beta}
\nonumber \\       
       &&+ \Delta T_{\rm shear}^{\alpha \beta}
       + \Delta T_{\rm cond}^{\alpha \beta},
\label{Tabab} \\
       \Delta T_{\rm shear}^{\alpha \beta} &=&
       - \eta \, H^{\alpha \gamma} H^{\beta \delta} \,
       \left( u_{\gamma \, ; \delta} + u_{\delta \, ; \gamma}
       - {2 \over 3} \,\, g_{\gamma \delta} \,
       u^{\lambda}_{\,\, ; \lambda} \right),
\label{shear} \\
       \Delta T_{\rm cond}^{\alpha \beta} &=& - \kappa \,
       (H^{\alpha \gamma} u^{\beta} + H^{\beta \gamma} u^{\alpha})
       \, (T_{\, ; \gamma} + T \, u_{\gamma \, ; \delta} \, u^{\delta}),
\label{cond}
\end{eqnarray}
where $g^{\alpha \beta}$ is the metric tensor, 
$\eta$ is the shear viscosity coefficient, 
$\kappa$ is the thermal 
conductivity, and 
$H^{\alpha \beta}= g^{\alpha \beta} + u^{\alpha} \, u^{\beta}$ 
is the projection matrix. 
In this paper we use $\eta = \eta_{\rm e}$,
where the electron shear viscosity $\eta_{\rm e}$ 
in the stellar core is taken from 
Chugunov $\&$ Yakovlev (2005). We neglect the shear 
viscosity of neutrons 
(the proton shear viscosity is even smaller, 
see Flowers $\&$ Itoh 1979) 
because it strongly depends on the nuclear interaction 
model and many-body theory employed. 
A similar problem for heat conduction was discussed 
by Baiko, Haensel $\&$ Yakovlev (2001).
The neutron shear viscosity is comparable 
to the electron shear viscosity (Flowers $\&$ Itoh 1979), 
but it cannot change our results qualitatively.

The use of Eqs.\ (\ref{consnb}), (\ref{consne}) and (\ref{Tab}) 
together with the second law of 
thermodynamics  
($\dd \rho = \mu_{\rm n} \dd n_{\rm n}
+ \mu_{\rm p} \dd n_{\rm p} + \mu_{\rm e} \dd n_{\rm e} + T \dd S$)
can yield the 
continuity equation for the entropy in the neutron star core
(see, e.g., Landau $\&$ Lifshitz 1959; Weinberg 1971):
\begin{eqnarray}
      (S u^{\alpha})_{; \alpha} &=&
      \left(
      Q_{\rm bulk}+Q_{\rm shear}+Q_{\rm cond} - Q_{\nu}
      \right)/k_{\rm B}T, 
\label{entropy} \\
   Q_{\rm bulk}&=& \delta \mu \, \Delta \Gamma, \quad
   Q_{\rm shear}=(\Delta T^{\alpha \beta}_{\rm shear})_{; \beta}
   \, u_\alpha, 
\label{QQQ} \\
   Q_{\rm cond}&=&( \Delta T^{\alpha \beta}_{\rm cond})_{; \beta}
   \, u_\alpha.
\label{Qcond}
\end{eqnarray}
Here $S$ is the entropy density and $Q_{\rm bulk}$ 
is the pulsation energy dissipating 
into heat per unit volume per unit time 
owing to the non-equilibrium modified 
Urca process. 
The latter term can be interpreted 
as viscous dissipation due to an
{\it effective bulk viscosity}. 
We will show below that at $\delta \mu \ll k_{\rm B}T$ 
it coincides with the term commonly considered 
by other authors
(see, e.g., Sawyer 1989 or Haensel et al.\ 2001). 
The term $Q_{\rm shear}$ describes 
the dissipation of pulsation 
energy into heat due to the shear viscosity. 
The term $Q_{\rm cond}$ is generally 
responsible for heat diffusion in the star bulk and for 
the dissipation of pulsation 
energy due to heat conduction. 
Finally, $Q_\nu$ 
is the neutrino emissivity. 
In this work, the quantity $Q_{\rm cond}$ 
was calculated using an unperturbed 
metric (the metric (\ref{ds}) 
with $\nu=\nu_0$ and $\lambda=\lambda_0$)
neglecting temperature variations 
over a pulsation period. 
The result coincides with the
similar expression well-known 
in the cooling theory
of non-pulsating neutron stars 
(see, e.g., Thorne 1977; van Riper 1991). 
These assumptions are quite reasonable in the case 
of a strongly degenerate matter. 
The damping due to heat conduction has been 
analyzed by Cutler $\&$ Lindblom (1987) 
for a more general case of non-radial pulsations. 
The conclusion made by these authors 
is that the contribution of 
heat conduction to the dissipation 
of pulsation energy can be neglected.

For the spherically symmetric metric ({\ref{ds}}), 
the left-hand side of Eq.\ (\ref{entropy}) 
can be rewritten as
\begin{align}
   &(S u^{\alpha})_{; \alpha} =
   \nonumber \\
   &=
   \left[ {\partial (S \,{\rm e}^{\lambda/2}) \over \partial t}
   + {1 \over r^2} \, {\partial \over \partial r}
   \left( r^2  S \,{\rm e}^{\lambda/2} \,
   {\partial \xi(r,t) \over \partial t }\right)
   \right] \,\, {\rm e}^{-(\lambda+\nu)/2}.
\label{entropy1}
\end{align}
In our further treatment, we will use 
the {\it isothermal approximation}, 
in which the redshifted internal temperature  
is taken to be constant over the star bulk: 
${\widetilde T}=T {\rm e}^{\rm \nu/2}$ = {\rm const}.
This approximation works well for a cooling star of the age 
$t \ga (10-50)$ yrs 
(see, e.g., Yakovlev et al.\ 2001; Yakovlev $\&$ Pethick 2004). 
The isothermal approximation 
considerably simplifies calculations 
without any significant loss of accuracy
(at least for the case of non-equilibrium modified
Urca processes).  
Using Eq.\ (\ref{entropy1}) and the 
integral form of Eq.\ (\ref{entropy}), and averaging over a 
pulsation period, we arrive at the thermal balance equation:
\begin{align}
     &{\dd E_{\rm th} \over \dd t}  \equiv 
     C_T \, {\dd \widetilde{T} \over \dd t}
     = -L_{\rm phot} - L_\nu + W_{\rm bulk} + W_{\rm shear},
     \label{entropy2} \\
     &C_T=\int \, c_T \, \dd V,
\label{C_T} \\
     &L_{\rm phot} = 4 \pi R^2 \, \sigma T_{\rm s}^4 \,\,
     {\rm e}^{\nu_0(R)}, \quad \, 
     L_\nu = \int \,
     \overline{Q}_{\nu} \,\,
     {\rm e}^{\nu_0} \, \, \dd V,
\label{phot_nu} \\
     &W_{\rm bulk} = \int \,
     \overline{Q}_{\rm bulk} \,\,
     {\rm e}^{\nu_0} \, \, \dd V, \quad
     W_{\rm shear} = \int \,
     \overline{Q}_{\rm shear} \,\,
     {\rm e}^{\nu_0} \, \, \dd V.
\label{bulk_shear}
\end{align}
Here, $L_{\rm phot}$ and $L_{\nu}$ 
are the redshifted photon and neutrino luminosities of the star; 
$W_{\rm bulk}$ and $W_{\rm shear}$ 
denote the heat released in the 
star per unit time owing to the bulk and shear viscosities, respectively; 
$\sigma$ is the 
Stefan-Boltzmann constant; 
$T_{\rm s}$ is the effective surface temperature 
(the relationship between the surface 
and internal temperatures is reviewed, e.g., 
by Yakovlev $\&$ Pethick 2004). 
The upper horizontal line denotes  
averaging over a pulsation period. 
In deriving Eq.\ (\ref{entropy2}), 
we neglected the terms of 
the order of 
$\sim T^2 \varepsilon^2$, 
as compared  to those of $\sim T^2$, 
in the left-hand side of the equality. 
The term $Q_{\rm cond}$ in Eq.\ (\ref{entropy}) 
leads to the appearance of $L_{\rm phot}$ in Eq.\ (\ref{entropy2}).
	
The rate of the heat release due to the shear viscosity is
\begin{align}
     &\overline{Q}_{\rm shear} =
     \nonumber \\
     &= {\eta \over 3} \,\, { {\rm e}^{-\nu_0} \over r^2} \,\, 
     \overline{ \left\{
     -2 r { \partial^2 \xi(r,t) \over \partial r \partial t}
     + 2 { \partial \xi(r,t) \over \partial t}
     + r { \partial \xi(r,t) \over \partial t}
     {\dd \nu_0 \over \dd r} \right\}^2}
\nonumber \\
     &= {\eta \over 6} \,\, \omega_k^2 C_k^2(t) \,\,
     { {\rm e}^{-\nu_0} \over r^2} \,\, \left\{
     -2 r { \dd \xi_k \over \dd r}
     + 2 \xi_k  + r \xi_k
     {\dd \nu_0 \over \dd r} \right\}^2.
\label{shear_emis}
\end{align}
Among the processes contributing to 
the neutrino emissivity $Q_\nu$, the 
only process, whose emissivity $Q_{\rm noneq}$ 
can vary dramatically over a pulsation period, 
is the modified Urca process
(we assume that the direct Urca process is forbidden). 
The expression for $\overline{Q}_{\rm noneq}$ 
is obtained from Eq.\ (\ref{Q}) by averaging over the 
pulsation period $P = 2 \pi / \omega_k$ 
with allowance for $y(r,t)=y_0 \, \cos(\omega_k t)$, 
where $y_0$ is a slowly varying function of time:
\begin{align}
    &\overline{Q}_{\rm noneq} = Q_{\rm eq} \,
    \left( 1 + {11010  \pi^2\, y_0^2 \over 11513}
    \right.
    \nonumber \\
    &+ \left. {8505  \pi^4\, y_0^4 \over 46052} 
    + {525 \pi^6\, y_0^6 \over 46052 }
    + {315 \pi^8\, y_0^8 \over 1473664 } \right),
\label{Qnoneq} \\
    &y_0 = {C_k(t) \over \pi^2 k_{\rm B} T} \, \,
    {\partial \delta \mu(n_{\rm b0}, x_{\rm e0})
    \over \partial n_{\rm b0}} \, \,
    n_{\rm b0} \, {{\rm e}^{\nu_0/2} \over {r^2}} \,
    {\partial \over \partial r} \left( r^2 {\rm e}^{-\nu_0/2} \,
    \xi_k \right).
\label{y0}
\end{align}
Similarly, Eqs.\ (\ref{dG}) and (\ref{connection}) 
result in the expression for the heating rate 
produced by the dissipation of the pulsation energy 
due to deviations from beta-equilibrium:
\begin{eqnarray}
     \overline{Q}_{\rm bulk} &=&
     {14680 \pi^2\over 11513 } \, Q_{\rm eq}
\nonumber \\
&\times&
     \, \left( {y_0^2 \over 2} + {567 \pi^2\, y_0^4 \over 2936 }
     + {105 \pi^4\, y_0^6 \over 5872 } + {21  \pi^6\, y_0^8 \over 46976}
     \right).
\label{dmudG}
\end{eqnarray}
Analogous expressions for the non-equilibrium direct Urca process 
are presented in the Appendix.

The quantities $\overline{Q}_{\rm noneq}(y_0)$ and
$\overline{Q}_{\rm bulk}(y_0)$ 
for the modified Urca process were calculated
numerically by Finzi \& Wolf (1968). Their results (their Fig.\ 1)
at $y_0 \la 1$ are correct only qualitatively (Haensel 1992),
although they become exact in the limit of $y_0 \gg 1$.   

Using Eqs.\ (\ref{Qnoneq}) and (\ref{dmudG}), 
one can easily find the neutrino emissivity and 
the viscous dissipation rate 
of the non-equilibrium modified Urca process for both 
subthermal or suprathermal pulsations 
(if they are small, i.e. $\varepsilon \ll 1$). 
The typical value $\overline{y}_0$ 
of the parameter $y_0$ in the stellar core 
can be estimated for the fundamental mode as:
\begin{equation}
     \overline{y}_0
     \sim 100 \,\varepsilon C_k(t) / T_9. 
\label{estimation_y0}
\end{equation}
%

Thus, we have derived the 
equation describing the thermal evolution 
of a pulsating neutron star. 
This equation depends on 
the current pulsation amplitude 
$C_k(t)$ and, hence, on the 
pulsation energy $E_{\rm puls}(t)$ 
(see Eq.\ (\ref{Epuls1})). 
Let us now derive the equation to describe the 
evolution of the pulsation energy. 
In principle, it can be obtained from the 
``pulsation equation''
(59) of Chandrasekhar (1964) 
by taking into account the dissipation terms and 
considering them as small perturbations 
(generally non-linear in $\xi_k$). 
We have performed this derivation, 
but here we will present a much simpler derivation  
following from energy conservation law. 
One should bear in mind that the 
pulsation energy dissipates due 
to the bulk and shear viscosities 
and is fully spent to heat the star. 
The corresponding terms have already been found 
for the thermal balance equation (\ref{entropy2}). 
The same terms, but with the opposite sign, 
should be valid for the damping equation
which can be thus presented in the form:
\begin{equation}
      {\dd E_{\rm puls} \over \dd t} = -W_{\rm bulk} - W_{\rm shear}.
\label{pulsations}
\end{equation}
The set of Eqs.\ (\ref{entropy2}) and (\ref{pulsations}) 
has to be solved to obtain self-consistent solutions for the 
pulsation amplitude $C_k$ and 
temperature ${\widetilde T}$ as a function of time. 

Similar equations were formulated, analyzed and solved numerically
by Finzi \& Wolf (1968) under some simplified assumptions.
In particular, the authors neglected the pulsational damping
due to the shear viscosity ($W_{\rm shear}=0$).
They used approximate expressions for $L_\nu$ and
$W_{\rm bulk}$ (see above) and neglected the effects
of General Relativity. In addition, they used simplified
models of neutron stars and  stellar oscillations.
However, their approach was quite sufficient to understand
the main effects of the non-equilibrium modified Urca
process on the thermal evolution of neutron stars and
damping of their vibrations. We extend this consideration
using the updated microphysics input with the proper treatment of the effects of
the shear viscosity and General Relativity.

We can generally write:
\begin{align}
    & C_T = 10^{39} \,a_C \, \widetilde{T}_9~~
    {\rm erg~K}^{-1}, \,\,\,
    L_{\nu 0} = 10^{40} \, a_L \, \widetilde{ T}_9^8 \, \,
    {\rm erg~s}^{-1}, 
    \nonumber \\
    &E_{\rm puls} = 10^{53} \, a_P\,  \omega_4^2 \, \varepsilon^2
    C^2_k~~{\rm erg}, \quad E_{\rm th}=C_T \widetilde{T}/2, 
\nonumber \\
    &
    L_\nu = L_{\nu 0} \,
    \left( 1+ a_1 \, \overline{y}_0^2 + a_2 \, \overline{y}_0^4+
            a_3 \, \overline{y}_0^6 + a_4 \, \overline{y}_0^8       \right),
    \nonumber \\
    &W_{\rm shear} = 10^{38} \, a_S \, \omega_4^2\,
    \overline{y}_0^2~~{\rm erg~s}^{-1}, 
\nonumber \\    
    & W_{\rm bulk} = 
    L_{\nu 0} \,
    \left( {2 \over 3}\, a_1 \, \overline{y}_0^2 + 
            {4 \over 3}\, a_2 \, \overline{y}_0^4+
             2 a_3 \, \overline{y}_0^6 + 
	    {8\over 3} \, a_4 \, \overline{y}_0^8       \right), \quad
	    \nonumber \\
	    &\overline{y}_0 \equiv 
	    10^2 \,  \varepsilon \, C_k /{\widetilde T}_9.
\label{Numerical}
\end{align}
Here ${\widetilde T}_9 = {\widetilde T}/(10^9 \, {\rm K})$; 
$L_{\nu 0}$ is the neutrino luminosity of a non-pulsating star; 
$a_C$, $a_L$, $a_P$, $a_S$, $a_1$,\ldots,$a_4$ 
are dimensionless factors which depend on a stellar model 
and on a pulsation mode. 
For our model of a neutron star with 
$M=1.4\,M_\odot$ (the equation of state of 
Heiselberg $\&$ Hjorth-Jensen 1999) 
and for the fundamental pulsation mode we have obtained 
$a_C=1.88$, $a_L=5.34$, $a_P=1.81$, $a_S=4.75$,
$a_1=9.18$, $a_2=20.0$, 
$a_3=15.0$, $a_4=3.61$.

%
%

\section{Analytical Solutions and Limiting Cases}
\label{limits}

Before presenting numerical solutions to Eqs.\ 
(\ref{entropy2}) and (\ref{pulsations}), 
let us point out general 
properties of the solutions and 
consider the limiting cases. 
Numerical values will be given for the fundamental mode  
and for the above model of the star with $M=1.4\,M_\odot$.

Equation (\ref{pulsations}) describes 
the damping of pulsations due to the bulk and 
shear viscosities. 
In the present problem, 
there are no instabilities that 
could amplify stellar pulsations. 
In contrast, the thermal balance equation (\ref{entropy2}) permits 
both the stellar cooling 
(at $L_\nu + L_{\rm phot} > W_{\rm bulk} + W_{\rm shear}$) 
and the heating due to the viscous dissipation of 
the pulsation energy 
(at $L_\nu + L_{\rm phot} < W_{\rm bulk} + W_{\rm shear}$).

One may expect qualitatively 
different solutions in the subthermal 
($\overline{y}_0 \ll 1$)
and suprathermal ($\overline{y}_0 \gg 1$) regimes. 
For the former, we have 
$E_{\rm puls} \ll E_{\rm th}$,
while for the latter 
$E_{\rm puls} \gg E_{\rm th}$.

\subsection{The modified Urca regime}
\label{murca}

A sufficiently hot star has 
$L_{\rm phot} \ll L_\nu$
and $W_{\rm shear} \ll W_{\rm bulk}$. 
Then the evolution of pulsations and the thermal evolution 
of the star are totally 
determined by the non-equilibrium modified Urca process.
The main features of this {\it modified Urca regime} were
analyzed by Finzi \& Wolf (1968). 
We present this analysis
using more accurate approach
(see above).

This regime is conveniently studied by analyzing
the evolution of $\widetilde{T}(t)$ and $\overline{y}_0(t)$.
Neglecting $L_{\rm phot}$ and $W_{\rm shear}$,
we can rewrite Eqs.\ (\ref{entropy2}) and (\ref{pulsations}) as:
\begin{eqnarray}
  && {2 E_{\rm th} \over \overline{y}_0} 
  \, {\dd  \overline{y}_0 \over \dd t} 
  = L_\nu - W_{\rm bulk} \, 
  \left( 1 +  { E_{\rm th} \over E_{\rm puls} }\right)=
  \widetilde{T}^8 \, A(\overline{y}_0),
\label{y_ev} \\
&&   {2 E_{\rm th} \over \widetilde{T}} 
\,\, {\dd  \widetilde{T} \over \dd t} 
= -L_\nu + W_{\rm bulk}= 
  \widetilde{T}^8\,B(\overline{y}_0),
\label{T_ev}
\end{eqnarray}    
where 
$E_{\rm th}/E_{\rm puls}=a_C/ (20 \,\overline{y}_0^2 \, 
a_P \, \omega_4^2)$;
the functions $A(\overline{y}_0)$ and $B(\overline{y}_0)$ are
independent of $\widetilde{T}$. Their
exact form is easily deduced from Eq.\ (\ref{Numerical}).    
One immediately has ${\rm d} \ln \overline{y}_0 / {\rm d}
\ln \widetilde{T} = A(\overline{y}_0)/B(\overline{y}_0)$, which
allows one (in principle) to obtain the relation between
$\overline{y}_0$ and $\widetilde{T}$ in an integral form. 

Equations (\ref{y_ev}) and (\ref{T_ev}) have two
special solutions. 

The first solution is obvious and refers to an ordinary
non-vibrating ($\overline{y}_0(t)\equiv 0$) cooling neutron star.
In this case
\begin{eqnarray}
  \widetilde{T}(t) &=& \widetilde{T}(0) / 
  \left(1 + 6 \beta_0 \widetilde{T}_9^6(0) \, t \right)^{1/6}, 
\label{case1} 
\end{eqnarray}
where $\beta_0 = a_L \, /(10^8 a_C)$. We have 
$\beta_0 \approx 1/(1.12~{\rm yr})$,
for our neutron star model.  

The second solution is realized 
(Finzi \& Wolf 1968) if the initial value
$\overline{y}_0(0)$ satisfies the equation 
\begin{equation}
   {L_{\nu} \over W_{\rm bulk}} = 1+
   {E_{\rm th} \over E_{\rm puls}}=
    1 + {a_C \over 20 \, \, \overline{y}_0^2 \, 
    a_P \, \omega_4^2},
\label{Condition}
\end{equation}
at which $A(\overline{y}_0)=0$ and
${\rm d}\overline{y}(t)/{\rm d}t = 0$.
In this case $\overline{y}_0(t)$ {\it remains
constant} during the modified Urca stage.
We will denote this
specific value of $\overline{y}_0$ by $\overline{y}_{0L}$;
it is equal to
$\overline{y}_{0 L} \approx 0.607$ for our
neutron star model.
In this limiting case
$\widetilde{T}(t)$ and $C_k(t)$
are easily obtained from Eq.\ (\ref{entropy2}):
\begin{eqnarray}
  \widetilde{T}(t) &=& \widetilde{T}(0) / 
  \left(1 + 6 \beta \widetilde{T}_9^6(0) \, t \right)^{1/6},
\label{regime1temp} \\
  C_k(t) &=& \widetilde{T}_9 (t) \, 
  \overline{y}_{0 L}/10^2 \varepsilon, 
\label{regime1ampl} 
\end{eqnarray}
where $\beta = a_L \, (L_\nu-W_{\rm bulk})/(10^8 a_C\,L_{\nu 0})$ 
($\approx 1/(3.05\,{\rm yr})$, for our model).
Thus, the internal stellar temperature $\widetilde{T}(t)$ and
the pulsation amplitude $C_k(t)$ simultaneously decrease with time,
leaving the suprathermality level constant, intermediate between the
subthermal and suprathermal pulsation regimes.
The decrease is power-law (non-exponential).

The thermal evolution of neutron stars in the two
limiting cases is remarkably similar. If the star was
born sufficiently hot ($\widetilde{T}(0) \ga 10^9$ K)
then in a few years after the birth the initial
temperature becomes forgotten. For the non-vibrating star
from Eq.\ (\ref{case1}) we have 
$\widetilde{T}_9^{(1)}(t)\approx (6 \beta_0\, t)^{-1/6}$,
while for the vibrating star from Eq.\ (\ref{regime1temp}) we have
$\widetilde{T}_9^{(2)}(t)\approx (6 \beta \, t)^{-1/6}$.
Thus, the vibrating star stays somewhat hotter,
$\widetilde{T}^{(2)}(t)/\widetilde{T}^{(1)}(t)=(\beta_0/\beta)^{1/6}$
($\approx 1.18$ for our model).

Once the two limiting solutions are obtained, 
all other solutions for the modified Urca regime become clear.
If $\overline{y}_0(0)>\overline{y}_{0L}$, then
$A(\overline{y}_0(0))<0$ and $\overline{y}_0(t)$ will
tend to $\overline{y}_{0L}$ from above. 
If $\overline{y}_0(0)<\overline{y}_{0L}$, then
$A(\overline{y}_0(0))>0$ and $\overline{y}_0(t)$ will
tend to $\overline{y}_{0L}$ from below.
After $\overline{y}_0(t)$ comes sufficiently close
to  $\overline{y}_{0L}$, the stellar evolution
is approximately described by the limiting
solution given by Eqs.\ (\ref{regime1temp})
and (\ref{regime1ampl}). Therefore, this limiting
solution describes the {\it universal asymptotic behavior of
all vibrating neutron stars}.

\subsection{The damping of oscillations by the shear viscosity}
\label{shearuk}

One may expect qualitatively 
different solutions for the damping due 
to the bulk viscosity 
($W_{\rm bulk} \gg W_{\rm shear}$, a hot star) 
and the shear viscosity 
($W_{\rm shear} \gg W_{\rm bulk}$, a cold star). 
Using Eqs.\ (\ref{Numerical}), 
it is possible to show that the temperature $T_{\rm visc}$ 
separating these two regimes 
(and obeying the condition 
$W_{\rm bulk} \sim W_{\rm shear}$) is approximately 
equal to $T_{\rm visc} \sim 7 \times 10^8/(1+\overline{y}_0^2)^{3/8}$~K.
For the regime of damping due to shear viscosity
($T \ll T_{\rm visc}$), 
Eq.\ (\ref{pulsations}) reduces to a linear 
equation for $C_k(t)$, 
irrespectively of the value of $\overline{y}_0$:
\begin{equation}
{\dd C_k(t) \over \dd t} =
      - {\alpha_{\rm shear} \over 2 {\widetilde T}_9^2} \, C_k(t),
\label{pulsations1a}
\end{equation}
where 
$\alpha_{\rm shear} \approx 
3 \times 10^{-11}$~s$^{-1} \sim 1/(1000$ yrs)
is a constant factor. 
The solution to this equation shows an 
exponential damping, 
which is independent of
$\overline{y}_0$:
\begin{equation}
    C_k(t)=C_k(t_0)\, 
    \exp \left(- {\alpha_{\rm shear} \over 2}\, \int_{t_0}^t
    {{\rm d} t'\over \widetilde{T}_9^{2}(t')} \right).
\label{sheardamp}
\end{equation}
%

\subsection{Subthermal Pulsations}
\label{subthermal}

In this case ($y_0 \ll 1$), Eq.\ (\ref{dmudG}) 
can be reduced to
\begin{align}
      &\overline{Q}_{\rm bulk} = 
      {14680 \pi^2 \over 11513 } \, Q_{\rm eq} \, {y_0^2 \over 2} =
      \nonumber \\
       &= \zeta \, \overline{\left[ {1 \over r^2} \, {\partial \over \partial r}
       \left(
       r^2 \, {\rm e}^{-\nu_0/2} \,
       {\partial \xi(r,t) \over \partial t} \right)
       \right]^2}
      = \zeta \, \overline{(u^{\alpha}_{; \alpha})^2},
\label{dmudG1} \\
       &\zeta = {14680 \over 11513 \pi^2} \,
       {Q_{\rm eq} \over (k_{\rm B} T)^2} \,
       {n_{\rm b0}^2 \over (\omega_k \, {\rm e}^{-\nu_0/2})^2} \,
       \left[ {\partial \delta \mu(n_{\rm b0}, x_{\rm e0})
       \over \partial n_{\rm b0} } \right]^2.
\label{zeta}
\end{align}
The quantity $\zeta$ can be treated 
as the bulk viscosity. 
Equation (\ref{zeta}) coincides with the 
corresponding expression of Sawyer (1989) 
and Haensel et al.\ (2001). 
If the temperature remains constant 
during the damping, 
Eqs.\ (\ref{Epuls1}) and (\ref{pulsations}) 
yield an exponential fall
of the pulsation amplitude $C_k(t)$, which is often 
discussed in literature 
(see, e.g., Cutler et al.\ 1990).

According to Eq.\ (\ref{Numerical}), 
the subthermal regime is characterized by 
$L_\nu \approx L_{\nu 0} \gg W_{\rm bulk}$. In this case 
pulsations do not affect the neutrino luminosity, 
and the energy dissipation due 
to the bulk viscosity cannot produce a considerable stellar heating. 
The dissipation due to the shear viscosity is also too weak, 
$W_{\rm shear} \ll L_{\nu 0}$.
For these reasons, 
the pulsations do not 
change significantly the thermal balance 
equation (\ref{entropy2}) 
and the thermal evolution of the star. 
At the neutrino cooling stage 
(when $L_{\nu 0} \gg L_{\rm phot}$, 
which happens at $t \la 10^5$ yrs) 
we get the well-known formula (\ref{case1}) 
for non-superfluid neutron stars that 
cool via the modified Urca process. It can be
rewritten as (see, e.g., Yakovlev \& Pethick 2004)
\begin{equation}
   t = C_T \widetilde{T}/(6 L_{\nu 0})
     \sim 1~{\rm yr}/\widetilde{T}_9^6.
\label{tcool}
\end{equation}
This value of $t$ can be considered as 
a characteristic cooling time $t_{\rm cool}$
of the star with the internal temperature $\widetilde{T}$. 
For a hot star with 
$W_{\rm bulk} \gg W_{\rm shear}$
($T \gg T_{\rm visc}$; the modified Urca regime), 
Eq.\ (\ref{pulsations}) gives the 
characteristic pulsation damping time
\begin{equation}
   t_{\rm puls} \sim E_{\rm puls}/W_{\rm bulk} \sim t_{\rm cool}.
\label{tpuls}
\end{equation}
Therefore, the internal temperature $\widetilde{T}$ 
and the typical disbalance of the 
chemical potentials $\overline{\delta \mu}$ 
decrease with approximately the same characteristic time 
$t_{\rm cool}$ (see, e.g., Yakovlev et al.\ 2001). 
The parameter
$\overline{y}_0 \propto \overline{\delta \mu} / \widetilde{T}$,
which describes the
``level'' of pulsations 
relative to the thermal ``level'',
should tend to the limiting value $\overline{y}_{0L}$
(Sect.\ \ref{murca}). 
The damping of pulsations 
obeys the power law (rather than 
exponential, 
as would be in the absence of
the thermal evolution). 
This is because the viscous damping rate 
strongly depends on temperature, 
$W_{\rm bulk} \propto \widetilde{T}^6$.
 
In a cooler star 
($\widetilde{T} \ll T_{\rm visc}$, 
$W_{\rm shear} \gg W_{\rm bulk}$), 
the damping of subthermal 
pulsations is due to the 
shear viscosity and occurs, according 
to Eq.\ (\ref{sheardamp}), 
more abruptly (exponentially), decreasing 
the pulsation level 
$\overline{y}_0$.

\subsection{Suprathermal Pulsations}  
\label{suprathermal}

In this case, the quantity $\overline{Q}_{\rm bulk}$
cannot be generally described by an expression 
of the type of Eq. (\ref{dmudG1}). 
Strictly speaking, we cannot introduce a bulk viscosity 
$\zeta$, but Eq. (\ref{dmudG}) adequately describes 
the rate of the pulsation energy 
dissipation due to the modified 
Urca process. 
Nevertheless, 
at least for radial suprathermal pulsations,
the quantity $\overline{Q}_{\rm bulk}$ 
can be formally calculated from 
Eq.\ (\ref{dmudG1}), as before, with the effective 
bulk viscosity  
$\zeta$ given by Eq.\ (\ref{zeta}) 
with an additional factor 
$\overline{Q}_{\rm bulk}/\overline{Q}_{\rm bulk}(\overline{y}_0 \to 0)$.
In the suprathermal regime, 
the effective bulk viscosity and the viscous dissipation 
rate appear to be much larger than 
in the subthermal regime, 
as was pointed out by 
Haensel, Levenfish $\&$ Yakovlev (2002). 
However, there is an omission in their 
Eqs.\ (13)--(15) 
for the effective bulk viscosity 
in the suprathermal regime: 
the authors should have 
introduced an additional factor $\sim (1+ \overline{y}_0^2)$. 
This does not change 
qualitatively their principal results. 
Nevertheless, we stress that while analyzing
the damping of pulsations, 
one should account for 
the thermal evolution of the star. 
Accordingly, in the suprathermal regime Eq.\ (16) 
of Haensel et al.\ (2002) 
gives the characteristic time of {\it non-exponential} 
damping.

The pulsation equation (\ref{pulsations}) 
in the suprathermal regime 
($E_{\rm puls} \gg E_{\rm th}$,
$\overline{y}_0 \gg 1$)
at $W_{\rm bulk} \gg
W_{\rm shear}$ 
($\widetilde{T} \gg T_{\rm visc}$; the modified Urca regime) 
can be rewritten as
\begin{equation}
{\dd C_k^2(t) \over \dd t} = - \alpha_{\rm bulk} \, C_k^8(t),
\label{pulsations1}
\end{equation}
where 
$\alpha_{\rm bulk}\approx 3 \times 10^4\, 
\varepsilon^6 / \omega_4^2$~s$^{-1}$ 
is a constant factor. Assuming $C_k(0)=1$ we get:
\begin{equation}
C_k(t) \approx (1+3 \alpha_{\rm bulk} t)^{-1/6}.
\label{Ck(t)} 
\end{equation}
This solution describes a slow (power law) 
fall of the pulsation amplitude $\sim t^{-1/6}$ 
with the characteristic 
time $1/(3 \alpha_{\rm bulk})$. 
In this regime, the stellar heating always 
dominates over the cooling, 
with the heating rate $W_{\rm bulk} \approx 8/3\, L_\nu 
\propto \overline{\delta \mu}^8$  nearly 
independent of temperature
(being determined by 
the disbalance of the chemical 
potentials $\overline{\delta \mu}$).
This result was obtained by Finzi \& Wolf (1968). 
The power law decrease of $C_k(t)$ 
is associated with a strong dependence 
of $W_{\rm bulk}$ on $\overline{\delta \mu}$
(which mimics the dependence on $T$ in the 
subthermal regime). 
The relative pulsation amplitude  
should 
decrease, and the star should evolve 
to the subthermal regime ($\overline{y}_0 \to \overline{y}_{0L}$;
Sect.\ \ref{murca}). 

In a rather cool star, the damping
of pulsations due to the shear viscosity
dominates over the damping
caused by the bulk viscosity 
($W_{\rm shear} \gg W_{\rm bulk}$, 
$\widetilde{T} \la T_{\rm visc}$). 
The shear viscous damping is exponential, 
according to Eq.\ (\ref{pulsations1a}), 
so that the star 
rapidly evolves to the subthermal regime.     

\section{Results}
\label{results}

Generally, the set of Eqs.\ 
(\ref{entropy2}) and (\ref{pulsations}) 
has no analytical solution, and we 
have to solve it numerically. 
We have modified the isothermal version of our 
cooling code (for details, see the review of Yakovlev et al.\ 1999) 
by including a block for solving the damping 
equation (\ref{pulsations}).
Our code calculates the stellar surface 
temperature $T_{\rm s}^{\infty}$ 
(redshifted for a distant observer), 
as a function of time $t$, as well as $C_k(t)$. 
All the computations presented in this section are for the 
fundamental mode of radial pulsations. 
Computations for higher modes will not 
lead to qualitatively different conclusions.

The left panel of Fig. 2 shows the thermal evolution paths 
of a neutron star ($M=1.4\,M_\odot$), 
which differ in the initial 
internal temperature $\widetilde{T}(0)=\widetilde{ T}_0$
and the initial relative 
amplitude of pulsations $\varepsilon$ 
(see Eq. (\ref{CC})). 
The right panel presents 
$C_k(t)$ curves 
for the same models. 
The dotted curve on the left panel shows the cooling of a 
non-pulsating star (in the isothermal approximation). 
The circle indicates the 
observations of the Vela pulsar. 
References to the original observations can be 
found in Gusakov et al.\ (2004).

\begin{figure*}
\setlength{\unitlength}{1mm}
\leavevmode
\vspace*{10pt}
\includegraphics[width=160mm,bb=18  145  544  419,clip]{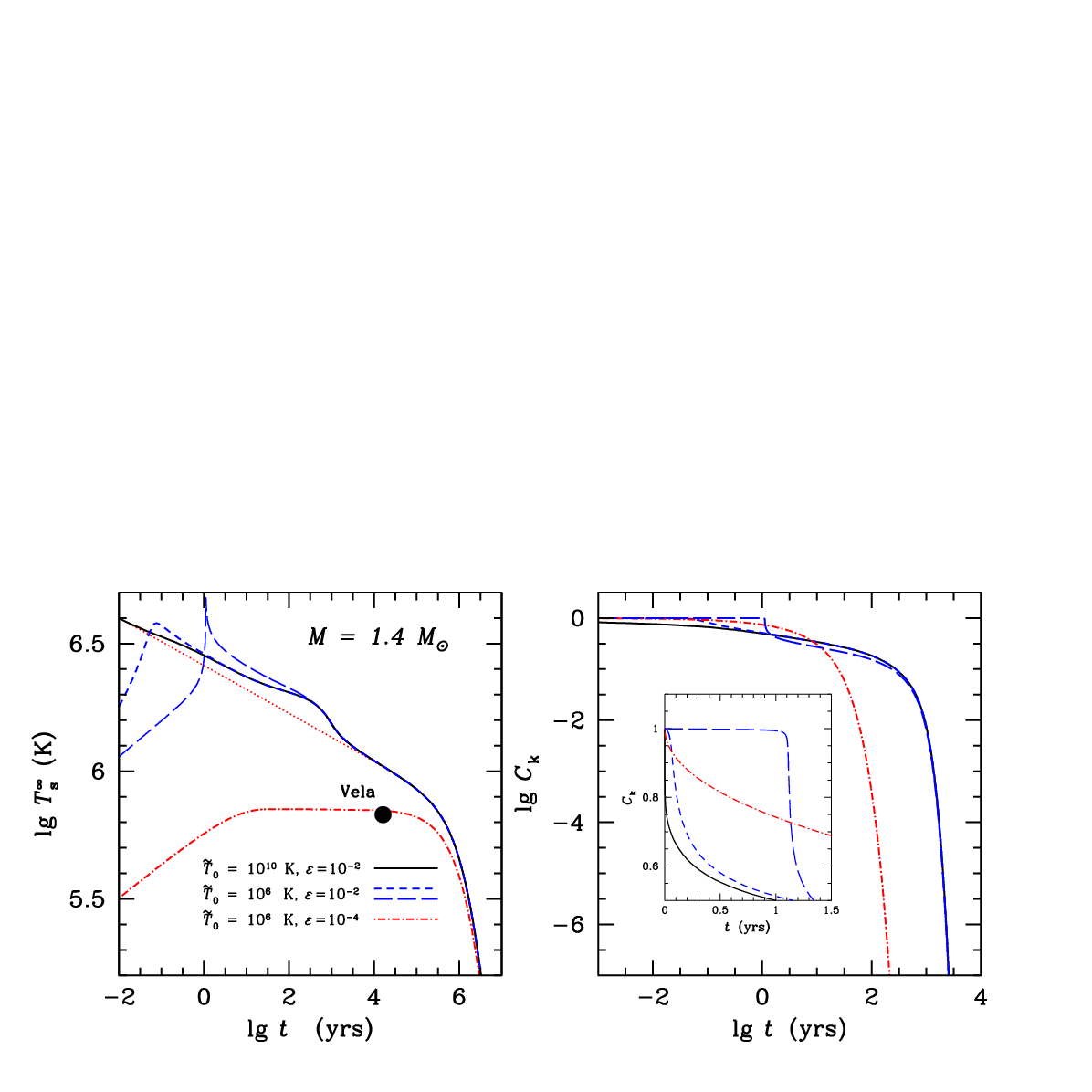}
\caption
{
Temporal evolution of the surface temperature (left) and pulsation 
amplitude (right) for three initial conditions: 
${\widetilde T}_0=10^{10}$ K, 
$\varepsilon = 0.01$ (solid lines); 
${\widetilde T}_0=10^{6}$ K, 
$\varepsilon = 0.01$ (dashed lines);
${\widetilde T}_0=10^{6}$ K, 
$\varepsilon = 0.0001$ (dash-dotted lines). 
The short dashed lines 
denote a self-consistent calculation, 
the long dashed lines are  
calculated neglecting non-linear effects 
due to deviations from the beta-equilibrium 
(see text). 
The dotted line on the left panel 
shows the cooling of a non-pulsating star. 
The full circle denotes the observations of the Vela pulsar. 
The insert displays the pulsation damping at $t \la 30$ yrs 
in more details.
}
\end{figure*}

The solid lines in both panels of Fig.\ 2 
are for the model with 
$\widetilde{T}_0=10^{10}$ K and $\varepsilon=0.01$.
This model describes a star which 
was born hot and strongly pulsating. 
The initial pulsation energy is about twice 
lower than its initial thermal energy, 
and the star is in an intermediate pulsation regime, 
between the supra- and subthermal regimes, 
with $\overline{\delta \mu}
\sim k_{\rm B} T$. 
The heating due to viscous dissipation is not as 
fast as the neutrino cooling due to 
the non-equilibrium modified Urca process, 
and the star is cooling down. 
The main contribution to the dissipation at the initial 
stage is produced by the bulk viscosity. 
The maximum difference between the surface 
temperatures of such star and 
a non-pulsating star occurs 
at $t \la 1000$ yrs. 
During this period of time,  
$\overline{\delta \mu}$ 
remains of the order of $k_{\rm B} T$ 
($\overline{y}_0 \approx \overline{y}_{0L}$).  
At $t \ga 1000$ yrs, 
the damping begins to be determined 
by the shear viscosity, which is not 
so temperature-dependent as the bulk viscosity. 
This leads to the exponential 
damping in the subthermal regime 
($E_{\rm puls}/E_{\rm th} \ll 1$; see the right panel of Fig.\ 2).

The short dashed lines in Fig.\ 2 
correspond to an initially cold star with 
$\widetilde {T}_0=10^6$~K and $\varepsilon=0.01$. 
The initial ratio of the pulsation energy to the thermal 
energy  is $E_{\rm puls0}/ E_{\rm th0} \sim 5 \times 10^7$, 
i.e., the star pulsates in a strongly suprathermal regime.  
As follows from Eq.\ (\ref{Numerical}), 
at low temperatures
we have $W_{\rm bulk} \ll W_{\rm shear}$, 
and the star is 
initially heated up by the shear viscosity. 
The heating due to the bulk viscosity 
starts to dominate only at 
$\widetilde{T} \ga 5 \times 10^8$ K. 
After heating to 
${\widetilde T} \approx 1.7 \times 10^9$ K
in $t \sim 1$ month, 
the star appears in the intermediate regime 
with $\overline{\delta \mu} \sim k_{\rm B} T$ and 
begins to cool down. 
At $t \ga 10$ yrs, 
the star starts evolving along the same 
``universal'' path 
($\overline{y}_0 \approx \overline{y}_{0L}$)
as in the first model.

The long dashed curves are obtained 
for the same initial conditions 
but in the ``naive'' approximation neglecting 
non-linear effects 
in non-equilibrium beta-processes. In particular, 
the neutrino luminosity is taken to be 
$L_\nu=L_{\nu 0}$
and the damping due to the bulk viscosity 
is determined by Eqs.\ (\ref{dmudG1}) and (\ref{zeta}). 
One can see that this approximation leads to 
qualitatively incorrect results. The viscous heating 
during the first year after the pulsation excitation 
is much slower than 
in the scenario with non-linear effects, 
and the star heats up slowly. 
The neutrino luminosity is also lower,
which enables the star to heat to higher temperatures. 
In fact, the slow damping due to the bulk viscosity does not 
decrease the pulsation amplitude during the first year.

The dash-and-dot lines in Fig.\ 2 
refer to the cold star with 
$\widetilde{T}_0=10^6$~K and $\varepsilon=0.0001$.
The initial ratio of the pulsation and thermal 
energies is  
$E_{\rm puls0}/E_{\rm th0} \sim 5 \times 10^3$, 
which means that the star is initially in the suprathermal regime. 
Nevertheless, the pulsation energy 
$E_{\rm puls0} \sim 5 \times 10^{45}$ erg 
is insufficient to heat the star to a temperature 
at which the damping is determined by the bulk viscosity. 
For this reason, the pulsation 
energy is damped by 
the shear viscosity. 
The damping of pulsations takes  
$\sim$100 years (see the right panel of Fig.\ 2). 
At $t \ga 100$ yrs the star cools via
photon emission from the surface. 
It is clear from the left panel that this model can, 
in principle, explain the surface 
temperature of a neutron star 
with the same thermal X-ray 
luminosity as the Vela pulsar 
but with different history. 
For example, it may be an old and 
cold isolated neutron star, 
in which radial pulsations have been excited. 
In $\sim$10 years after 
the excitation, the star 
will be heated to the temperature of the Vela pulsar. 
In 100 years, the pulsations 
will die out but the star 
will stay warm for $\sim 10^5$ years 
before it starts cooling down noticeably. 
It should be emphasized that these results 
will not change if we take 
a lower initial temperature, 
e.g., $\widetilde{T}_0= 10^4$~K. 
This star will also acquire the surface 
temperature  
$T_{\rm s}^\infty \sim 7 \times 10^5$ K 
in a year and will emit in soft X-rays.
\begin{figure*}
\setlength{\unitlength}{1mm}
\leavevmode
\hskip  0mm
\includegraphics[width=120mm,bb=18  145  554  675,clip]{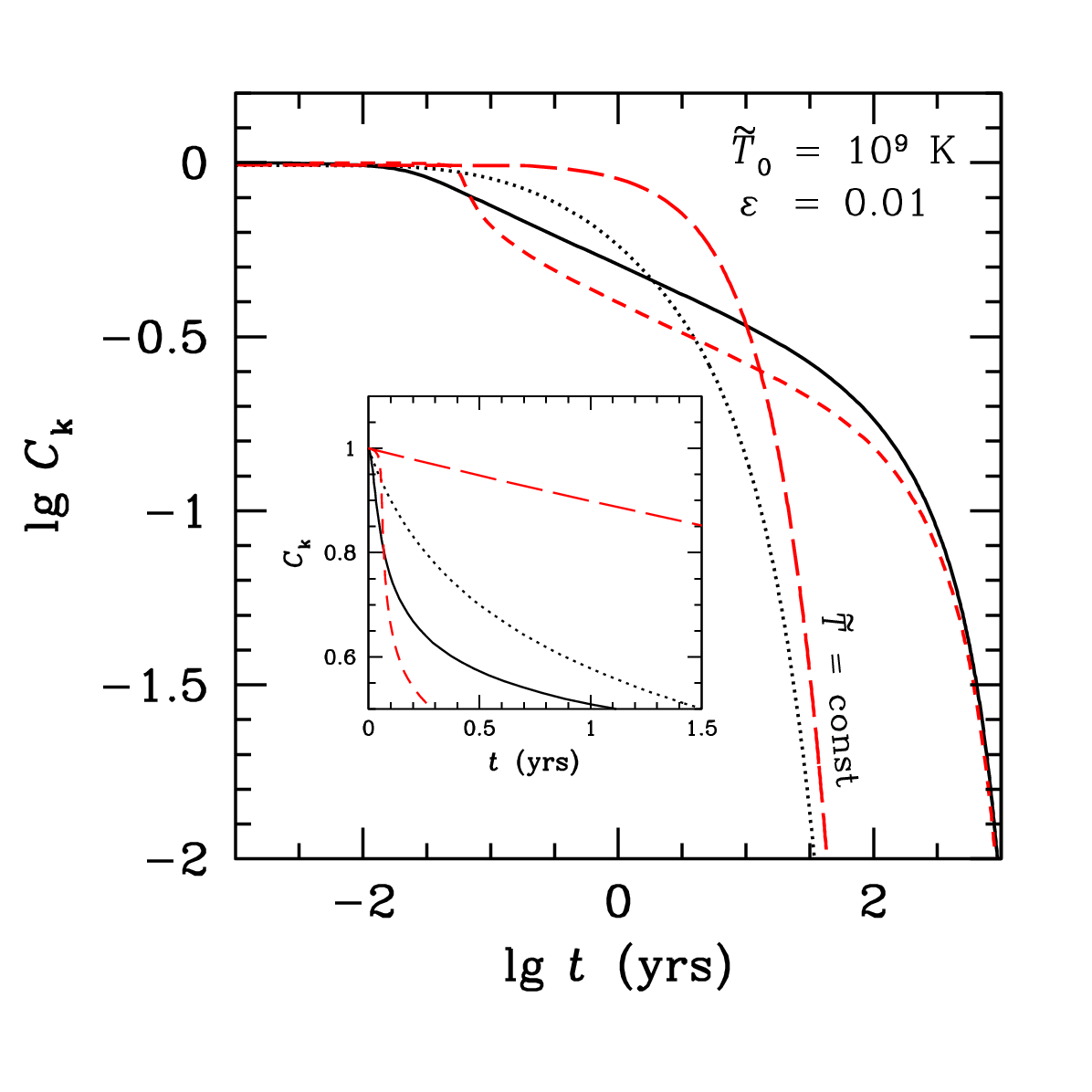}
\caption
{ 
Temporal evolution of the pulsation amplitude 
for the 
initial conditions: 
$\widetilde{T}_0=10^9$~K, $\varepsilon =0.01$.
Solid line presents a consistent calculation 
with account for the thermal evolution 
and non-linear deviations from beta-equilibrium (see text); 
dotted line shows the same but the internal
stellar temperature is fixed; 
short dashed line is the same as the solid line 
but without non-linear effects; 
long dashed line is obtained neglecting non-linear 
effects at the fixed temperature.
The insert displays the pulsation damping at $t \la 30$ yrs 
in more details.
}
\label{3}   
\end{figure*}

For a correct calculation of the pulsation damping, 
one should take into account the thermal evolution of the star. 
The evolutionary effects are especially important 
when the damping is determined 
by non-equilibrium beta-processes. 
They are relatively weak only in the subthermal regime, 
provided the damping is produced by shear viscosity.

These statements are also illustrated 
in Fig.\ 3 which shows the pulsation damping 
for a star with the initial internal 
temperature $\widetilde{T}(0)=10^9$~K 
and the initial relative pulsation 
amplitude $\varepsilon=0.01$.  
The initial ratio of the 
pulsation-to-thermal energy is
$E_{\rm puls0}/ E_{\rm th0} \sim 50$, 
indicating that the star is pulsating 
in a slightly suprathermal regime. 
The solid line is the result of 
a self-consistent solution of 
the thermal evolution and 
damping equations. 
The damping is power law 
for about 100 
years; afterwards the damping
is determined by the shear viscosity and becomes 
exponential. 
The pulsations die out 
completely in $\sim$1000 years.

The dotted line in Fig.\ 3 shows the solution 
to the damping equation neglecting 
the thermal evolution, 
at a constant internal temperature 
$\widetilde{T}=\widetilde{T}(0)$.
In this case, the 
pulsations are first damped by 
the bulk viscosity and then by
the shear viscosity in $\sim 30$ years. 

The short dashed curve is obtained by taking into account 
the thermal evolution and damping, 
but neglecting the non-linear 
effects in non-equilibrium beta-processes. 
For about 100 years, 
the damping is governed by the 
bulk viscosity; it is power law, 
but slower than with the non-linear effects. 
Later, the shear viscosity becomes important,
leading to the exponential damping, 
nearly the same as with the 
non-linear effects. 

Finally, the long dashed curve 
is calculated neglecting 
both the thermal evolution 
and the non-linear effects. 
Like in the case with these 
effects (the dotted curve), 
the damping is steep (exponential), 
taking about 
30 years, but occurs slightly slower
(the long dashed curve is above the 
dotted curve).

\section{Summary}
\label{summary}

Extending the consideration of Finzi \& Wolf (1968) 
we have analyzed the thermal evolution 
of a non-superfluid star which undergoes 
small-amplitude radial pulsations. 
We have derived a set of equations 
to describe the thermal evolution 
and the damping of pulsations 
in the frame of General 
Relativity. We have included the effects of 
non-linear deviations from beta-equilibrium 
in the modified Urca process
on the neutrino luminosity 
and on the pulsation energy dissipation due to the bulk 
viscosity in the stellar core. 
We have also taken into account
the dissipation due to the shear viscosity 
and the associated heating. 
A set of equations for the evolution of a neutron star with a 
nucleon core, in which the direct Urca process 
is forbidden, has been analyzed and solved analytically 
and numerically.

We have shown that the evolution of 
a pulsating star strongly depends on the 
degree of non-linearity of 
the non-equilibrium modified Urca process and on  
the nature of the pulsation damping  
(the shear or bulk viscosity). 
In the non-linear regime, the star may be 
considerably heated by 
the pulsation energy dissipation but 
it is always cooled down in the linear regime. 
Pulsations of a hot star are 
damped by the bulk viscosity 
in both the linear and non-linear regimes, 
and this process is rather slow 
(power law). 
In a cooler star, the 
damping is produced by the shear viscosity 
and goes much faster (exponentially). 
The characteristic times of damping
of the fundamental mode lie within 
100--1000 years.

We have not discussed here 
the specific damping mechanism via the 
ambipolar diffusion of electrons and protons 
relative to neutrons
when the averaged (over the period) 
chemical composition of the stellar matter 
remains constant in time.
As far as we know, this 
mechanism of the pulsation damping  
has not been analyzed 
in the literature. 
However, it may be as efficient 
as the damping by the shear viscosity, at 
least in the suprathermal regime. 
We will consider this problem in a separate
publication.

The analysis presented here 
is based on a simplified model. 
In particular, if  
the direct Urca process is open in the stellar core 
or if the core contains hyperons or quarks, 
the bulk viscosity can be many orders 
of magnitude higher than discussed here 
(see, e.g., Haensel et al.\ 2002 and references therein). 
The results may also differ significantly 
for superfluid neutron stars, because superfluidity 
drastically changes the reaction rates 
in dense matter and, hence, its kinetic 
properties, including the viscosity. 
It would also be instructive 
to consider other types of 
neutron star pulsations, 
primarily r-modes. 
They can be accompanied by the emission of 
gravitational waves 
(see, e.g., Andersson $\&$ Kokkotas 2001) 
which can, in principle, be registered by  
gravitational detectors of new generation. 
We expect to continue the analysis 
of the evolution of pulsating neutron stars. 

\section*{Acknowledgments}

We are very grateful to the anonymous referee
for pointing out the paper by Finzi \& Wolf (1968).
We are also grateful to D.P.\ Barsukov 
for numerous discussions; to A.I.\ Tsygan 
for constructive criticism; 
to N.\ Andersson, P.\ Haensel, A.D.\ Kaminker 
and K.P.\ Levenfish, 
for their interest in our work 
and valuable comments. 
One of the authors (MEG) 
also acknowledges  excellent 
working conditions at the 
N.\ Copernicus Astronomical Center in Warsaw, 
where this study was completed.

This research was supported 
by RFBR (grants 05-02-16245 and 03-07-90200), 
the Russian Leading Science School (grant 1115.2003.2), 
INTAS YSF (grant 03-55-2397), and
by the Russian Science Support Foundation.

\appendix

\section[]{Non-equilibrium Direct Urca Process}
\label{appendix}

If the non-equilibrium direct Urca process 
is allowed in a pulsating neutron star, 
it can also  be described by the quantities $Q_{\rm noneq}$ 
and $Q_{\rm bulk}$ given by Eqs.\ (\ref{Q}) and (\ref{QQQ}), respectively. 
These quantities are taken from 
Yakovlev et al.\ (2001) 
and are denoted here 
as $Q_{\rm noneq}^{(\rm D)}(y)$ and $Q_{\rm bulk}^{(\rm D)}(y)$.
The averaging over a pulsation period yields
\begin{eqnarray}
   {\overline Q}_{\rm noneq}^{(\rm D)} &=& Q_{\rm eq}^{(\rm D)} \,\,
   \left( 1 + {1071 \pi^2 \, y_0^2 \over 914} \right.
   \nonumber \\
   &+& \left.{945 \pi^4 \, y_0^4 \over 3656} +
   {105 \pi^6 \, y_0^6 \over 7312} \right) \, ,
\label{QDnoneq} \\
   {\overline Q}_{\rm bulk}^{(\rm D)} &=&
   {714 \pi^2 \over 457} \, Q_{\rm eq}^{(\rm D)} \,\,
   \left(
   { y_0^2 \over 2} +
   {15 \pi^2 \, y_0^4  \over 68}+
   {5 \pi^4 \, y_0^6 \over 272}
   \right) \,,
\label{QDbulk}
\end{eqnarray}
where $Q_{\rm eq}^{(\rm D)}$ 
is the neutrino emissivity 
of the direct Urca process 
and (as before) $y_0=\delta \mu/(\pi^2 k_{\rm B} T)$. 
In a pulsating star 
with the allowed direct Urca process, 
${\overline Q}_{\rm noneq}^{(\rm D)}$
and ${\overline Q}_{\rm bulk}^{(\rm D)}$ 
should be included 
into the quantities ${\overline Q}_{\rm noneq}$ 
and ${\overline Q}_{\rm bulk}$
given by Eqs.\ (\ref{entropy2}) and (\ref{pulsations}),
respectively. 
In the absence of nucleon superfluidity, 
the contribution of the direct Urca 
process into ${\overline Q}_{\rm noneq}^{(\rm D)}$ 
and ${\overline Q}_{\rm bulk}^{(\rm D)}$
is 5 -- 7 orders of magnitude greater than that of the 
modified Urca process.   

\bsp

\label{lastpage}

\end{document}